\documentclass[aps,pra,twocolumn,showpacs,groupedaddress,citeautoscript]{revtex4}  
\usepackage{graphicx}  
\usepackage{dcolumn}   
\usepackage{bm}        
\usepackage{amssymb}   
\usepackage{graphicx}
\usepackage{color}
\usepackage{bm}
\usepackage{amssymb}
\usepackage{amsmath}
\usepackage{soul}
\usepackage{tikz}
\usetikzlibrary{shapes,arrows}
\usepackage{amsmath}
\usepackage{appendix}
\usepackage{multirow}
\usepackage[normalem]{ulem}
\usepackage{booktabs, psfrag, soul}

\newcommand*{\citen}[1]{%
  \begingroup
    \romannumeral-`\x 
    \setcitestyle{numbers}%
    \cite{#1}%
  \endgroup   
}

\begin{document}

\title{Interaction-induced persistent current enhancement in frustrated bosonic systems}
\author{Vipin Kerala Varma} \affiliation{The \textit{Abdus Salam} International Centre for Theoretical Physics, Trieste, Italy}
\author{Rams\'{e}s J. S\'{a}nchez} \affiliation{Bethe Center for Theoretical Physics, Universit\"{a}t Bonn, Germany}

%

\date{\today}

\vspace*{-1cm}

\begin{abstract}
We investigate the effect of next nearest-neighbour hopping on the zero temperature Drude weight or superfluidity in mesoscopic one-dimensional 
systems of (a) single particle with quasidisorder (Aubry-Andr\'{e} model) and (b) hardcore bosons with nearest-neighbour interaction. 
We show that there is an interaction-induced enhancement of the Drude weight when the next nearest-neighbour hopping is frustrated for 
the many-body system. The observed non-monotonic behaviour of the Drude weight occurs because the repulsive interactions first suppress 
the frustration in the system, leading to a rise in the Drude weight, whereas at much larger interaction strengths the charge density 
wave fluctuations set in the insulator and the Drude weight drops. The present work reveals a novel scenario in which a persistent flow 
enhancement is plausible in the presence of kinetic frustration and repulsive interactions.
\end{abstract}

\maketitle
\section{Introduction}
An important feature distinguishing quantum mechanical systems from classical systems is the Aharonov-Bohm effect \cite{Bohm}{}, 
wherein a charged particle picks up a phase while moving through a field-free region which encloses a finite magnetic flux. Such a 
flux is proportional to the vector potential along the path of the particle and perturbs the energy levels of the quantum 
mechanical system \cite{Kohn}{}. A notable example of this effect occurs in low temperature mesoscopic rings in the presence of a 
static magnetic field, where the electronic wave function can retain the Aharonov-Bohm phase acquired throughout the ring and coherently 
extend over the whole system, thereby generating a persistent current \cite{Buettiker, Imry}{}. This is true even if the material is not 
superconducting, as had been predicted by theoretical studies and
corroborated by experiments on copper \cite{Levy} and gold \cite{Kleinsasser} 
rings, although these persistent currents are orders of magnitude smaller than those in superconductors \cite{Imry}{}. Remarkably 
persistent currents can also be generated in systems of neutral particles, 
where an artificial gauge field \cite{Dalibard, Sengstock, Spielman} or a rotation of the system causes the particles to acquire an 
Aharonov-Bohm phase. Indeed, a persistent flow has been achieved in an ultracold setup of Bose-Einstein condensed (BEC) atoms trapped 
in a toroidal geometry, by transferring orbital angular momentum to the atoms of the system \cite{Ryu}{}. 

A primary question in past studies, prompted mainly by the disagreement between experimental findings in small metallic rings and 
single-particle calculations, has been to discern the influence of interactions and disorder on the persistent currents in one- and 
two-dimensional model systems, see e.g. Refs. \citen{Giamarchi, Berkovits, Sarma, Poilblanc, BerkovitsII} and references therein. It is 
generally understood that repulsive interactions can counteract the localising effects of disorder, thereby increasing the persistent 
current in the system \cite{Giamarchi, Berkovits, Sarma}{}. However, such an interaction-induced enhancement of the current is by no
means predicted in all model systems \cite{Poilblanc}{}, or it is expected only to a weak extent \cite{BerkovitsII, Abrahams}{}. With 
the observation of a BEC superflow and the ability to artificially gauge model Hamiltonians, ultracold atomic setups seem ideal to 
investigate the effects of interactions and disorder on the persistent current in various physical models. 

In this work we investigate the zero-temperature Drude weight or conductivity stiffness (which determines the persistent response to an 
external field, and which is precisely defined below) in two prototypical systems that have been realised in cold atom experiments: 
single-particle model with incommensurate potential (Aubry-Andr\'{e} model) \cite{Inguscio} and hardcore bosons model 
\cite{Bloch, Simon, Lukin}{} at and away from half-filling. In particular, we ask what the effect of a longer-range hopping of the 
particles will be on the 
properties of each of these models defined on ring geometries. Our motivation for this specific question is two-fold: (i) the capability 
to generate artificial gauge fields \cite{Dalibard, Sengstock, Spielman}{}, persistent flows \cite{Ryu}{}, and tunable nearest- and next 
nearest-neighbour hopping amplitudes \cite{Struck} in ultracold systems raises the question of how longer range interactions and hopping 
processes of the gauged particles will affect the experimentally observable 
results; (ii) ladder materials, to which our long-range hopping one-dimensional systems are equivalent, realise rich physics through a 
variety of spin models\cite{Dagotto}{}. Moreover, their bosonic analogues have been studied in the literature and novel phases such as 
a supersolid have been recently predicted \cite{Mukerjee, Mishra}{}.

We now briefly summarise previous related work and results. Although the problem of a longer-range hopping has been addressed before for the 
aforementioned models \cite{Cesari, Biddle, Mukerjee, Ramakumar, Ogata}{}, these works focused largely on critical, gap and localisation properties. 
We will uncover some novel aspects missed in these earlier studies, namely that the interplay between kinetic frustration and non-local 
repulsive interactions produces a non-monotonic behaviour of the Drude weight as the interaction strength is increased. Our results will 
bear similarities to earlier reports on frustration-induced metallicity \cite{Ogata, Ogata2}{}, undertaken in the context of analysing 
the phases of the transition metal oxide $\textrm{PrBa}_2\textrm{Cu}_4\textrm{O}_8$ and of two-dimensional organic conductors. In these 
studies (i) the Ising gap was seen to vanish for a range of frustrated interaction strengths in a spin-system on a ladder 
geometry \cite{Ogata} and such a behaviour was interpreted as the onset of a conducting phase due to frustration. However, no analysis 
of the conductivity within the metallic phase was attempted; (ii) the Drude weight was seen to be enhanced by frustrated interactions 
in a quarter-filled two-dimensional electronic system \cite{Ogata2}{}. Our work extends and complements these studies by considering a 
kinetically-frustrated bosonic system where we find the Drude weight or superfluid density (and hence the persistent flow) 
to be enhanced by repulsive interactions, even in the thermodynamic limit;
that is, \textit{in the presence of frustration interactions can indeed enhance conductivity}.
We also explicitly demonstrate how such an effect is absent both for a single-particle one-dimensional system which undergoes a genuine conductor-insulator transition, and 
for the many-body system in the absence of frustration. A recent study \cite{Cominotti} of interacting one-dimensional bosons in the continuum with a localised and moving barrier 
in a ring reached a qualitatively similar conclusion: that there is an optimal on-site interaction for which the persistent flow is 
maximal.

\vspace*{-0.5cm}
\subsection*{Drude weight and superfluid density}
The persistent flow in a normal metal, superconductor or superfluid may be induced by threading a flux through a toroidal or ring 
system. Here we consider a ring of $L$ sites threaded by a flux $\phi$. This flux gives rise to a static vector potential along the 
path of the particles, which modifies the hopping amplitudes by a Peierls phase factor \cite{Peierls}{} 
($t \rightarrow t e^{\pm \phi/L}$) and perturbs the energy levels of the quantum mechanical system \cite{Kohn}{}. This 
sensitivity of the ground-state energy $E$ to the external flux produces a thermodynamic persistent flow $I \propto -dE/d\phi$ 
around the ring \cite{F_Bloch, Imry}{}. 

The response of the ground-state energy to an infinitesimal flux is measured by the Drude weight $\pi D$, which quantifies the strength 
of the zero-frequency peak in the real part of the Kubo conductivity. In one-dimensional systems it is given by \cite{Kohn, Shastry, Scalapino, Fisher} 

\begin{equation}
 \label{eq: SFdensity}
 D =  \frac{L}{2}\frac{d^2 E(\phi)}{d\phi^2} \biggr |_{\phi=\phi_0} \approx \frac{L \bigr(E(\phi) - E(\phi_0)\bigr)}{(\phi-\phi_0)^2} \equiv \frac{1}{2}\left(\frac{\rho}{m}\right)^*, 
\end{equation}

\noindent
where $\phi_0$ is the flux at which the ground-state energy is minimum and $\left(\rho/m\right)^*$ is the effective ratio of the 
density of mobile carriers to mass. Thus, for an infinitesimal flux, the Drude weight provides a way of quantifying the persistent 
flow in the ring \cite{Giamarchi}{}. Equivalently, the Drude weight can be understood as a measure of the sensitivity of the ground 
state to a small twist $\phi$ in the boundary conditions. Consequently, a vanishing (finite) Drude weight signals an insulating 
(conducting) phase at zero temperature \cite{Kohn}{}. Finally, we point out that the superfluid density is defined very similarly as 
the Drude weight in eq.~\eqref{eq: SFdensity}{}. Specifically, it corresponds to calculating first the ground-state energy in the 
thermodynamic limit and then its second derivative with respect to the flux \cite{Scalapino}{}. Nonetheless, in (quasi) 
one-dimensional systems $D$ and the superfluid density are equal due to the finite number of energy level crossings in the thermodynamic 
limit \cite{Scalapino, Hayward}{}. Therefore, our results for the Drude weight may also be interpreted as a measure of the superfluid 
density in the system. 

In what follows we compute, using exact diagonalisation, the ground-state energy of the two aforementioned models 
(both with and without a flux) defined on chains of length $L$ with periodic boundary conditions. The largest matrix we diagonalise 
has a linear dimension of about $10$ millon.

\begin{figure}[ttp!]
\centering 
\includegraphics[width=9cm]{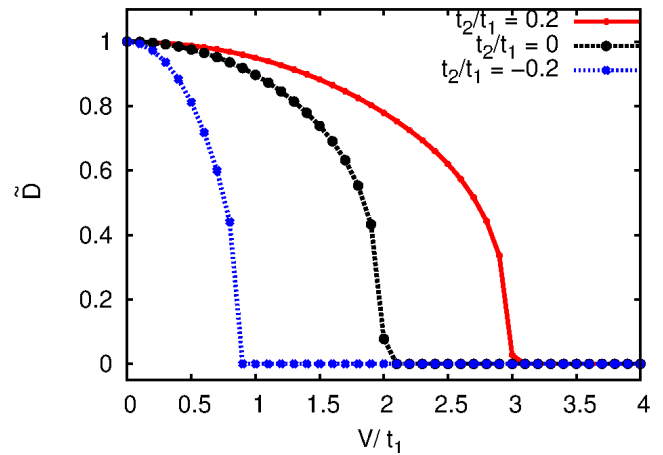}
\caption{(Colour online) Renormalised Drude weight $\tilde{D}$ (see text) in extended Aubry-Andr\'{e} model as a function of the 
quasi-commensurate potential strength $V$.}
\label{fig: AubryAndre}
\end{figure}

\vspace*{-0.25cm}
\section{Single particle system}
In this section we extend the results presented in Ref. \citen{Cesari}{}, as regards the superfluid density or Drude weight of the 
Aubry-Andr\'e model \cite{Aubry}{}, to the case with next nearest-neighbour (NNN) hopping. The model, defined 
on a ring of $L$ sites, reads

\begin{eqnarray}
 H &=& -t_1\sum_{j}^L (b_j^{\dagger}b_{j+1}^{\phantom{\dagger}} + \textrm{h.c}) -t_2\sum_{j}^L
 (b_j^{\dagger}b_{j+2}^{\phantom{\dagger}} + \textrm{h.c})\nonumber \\
& & + \sum_{j}^L \epsilon_j \hat{n}_j,
\raggedleft
\label{hamiltonian}
\end{eqnarray}

\noindent
where $j$ runs over the lattice sites, $b$ ($b^{\dagger}$) is the standard annihilation (creation) operator and $\hat{n}$ is the 
corresponding particle number operator. The hopping amplitudes to the nearest and next nearest neighbours are set respectively by 
$t_1$ and $t_2$, and $\epsilon_j=V \cos(2\pi j g)$ with $V$ the potential strength and $g = (\sqrt{5} + 1)/2$ the golden ratio. We 
note that the sign of $t_1$ in eq.~\eqref{hamiltonian} does not have any effect on the physics of the model. On the other hand, a 
negative value of $t_2$ does not allow the energy associated with nearest and next-nearest neighbour hopping processes to be 
simultaneously minimized i.e. it yields kinetic frustration.

For a single particle with no incommensurate potential ($V = 0$), eq.~\eqref{eq: SFdensity} may be readily evaluated to give 
$D = (t_1 + 4t_2)/L \equiv t_{\textrm{eff}}/L = \rho/2m$, where $\rho$ and $m$ are the system's density and 
the particle's bare mass. 
In order to facilitate comparison of results between different $t_2$ 
values, we define a renormalised Drude weight $\tilde{D} := L D/(t_{\textrm{eff}})$ which yields $\tilde{D} = 1$ for a single particle 
in the absence of disorder for any value of $t_2$. 

We consider a chain length of Fibonacci number $L=610$ in order to impose periodic boundary conditions \cite{Cesari}{}, and compute 
the Drude weight for the single-particle Aubry-Andr\'{e} model with NNN hopping amplitudes $t_2 = 0, \pm 0.2  \, t_1$. Our results 
are shown in Fig.~\ref{fig: AubryAndre} and bear out the following expectations: (i) there is a genuine conductor-insulator transition 
as a function of the potential strength $V$ in the ground state, as seen by the sharp drop in the Drude weights. Indeed, the transition 
point $V=2 \, t_1$ is known exactly for the nearest-neighbour Aubry-Andr\'{e} model; (ii) a negative value of the hopping $t_2$ yields 
a reduced $t_{\textrm{eff}}$ and works as a localising mechanism. Consequently, a weaker quasi-disorder is sufficient to fully localise 
the wave function. In contrast, the critical disorder strength increases for a positive $t_2$. These results are in quantitative 
agreement with inverse participation ratio (IPR) and Shannon entropy calculations \cite{Ramakumar}{}. However, we note that through our 
Drude weight calculations the conductor-insulator transition in the ground state may be pinpointed more accurately than through IPR 
calculations. 
The primary result of this section $-$ 
that the localising effect of the incommensurate potential always produces a monotonic decay of the Drude weight, although much 
slower for the unfrustrated system 
$-$ will be contrasted with the delocalising effect of the non-local repulsive 
interaction (and its corresponding effect on the persistent flow) while competing with kinetic frustration: in 
particular, the competition between interaction and frustration will be shown to produce a non-monotonic decrease of the 
Drude weight before the conductor-insulator transition sets in.

\begin{figure} [bbp!]
 \centering
 \includegraphics[width=9cm]{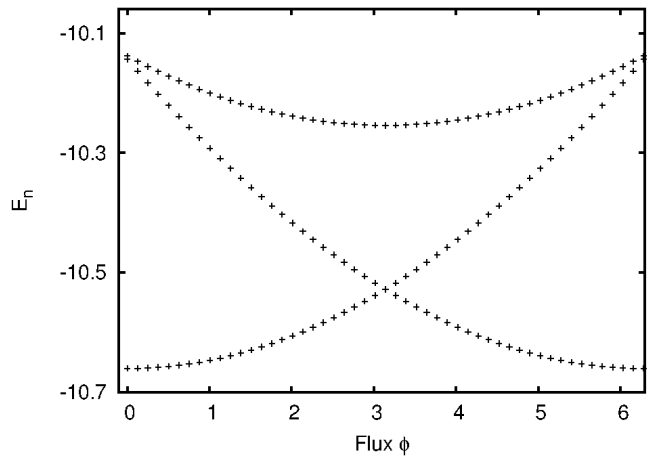}
  \caption{Energy levels as a function of flux in a 16-site ring with 2 hardcore bosons and system parameters 
  $t_2/t_1 = 0.2, V/t_1 = 1$. }
 \label{fig: Energy}
 \end{figure}
 \vspace*{-0.25cm}
\section{Hardcore boson systems}
Next we consider a system of hardcore bosons described by the Hamiltonian
\begin{eqnarray}
\label{eq: HCB}
H_{hcb} &=& H + V\sum_{i}\hat{n}_i\hat{n}_{i+1},
\raggedleft
\label{hamiltonian2}
\end{eqnarray}
where $V$ now sets the strength of nearest-neighbour interaction, $\epsilon_j = 0$ and the particle operators satisfy the usual 
hardcore boson commutation relations. Such a model has been extensively investigated recently \cite{Mukerjee, Mishra} in the context 
of analysing the competition between quantum fluctuations and kinetic frustration induced by the $t_2$ hopping term, and whether such a 
scenario can lead to exotic phases. We will work along a contour of the phase diagram presented by Mishra \textit{et al.} 
(Fig. 2 in Ref. \citen{Mukerjee}) and uncover some unique properties of the system in the conducting phase. 

Given that the second derivative in eq.~\eqref{eq: SFdensity} is evaluated at the location $\phi_0$ where the ground-state energy 
$E(\phi)$ is minimum, the dependence of the many-body energy levels on the flux deserves some explanation. In the absence of next 
nearest-neighbour hopping, the hardcore boson model is exactly solvable and the Drude weight may be written in closed form 
\cite{Shastry}{}. The model may also be Jordan-Wigner transformed to a system of spinless electrons and most results from the 
bosonic model transfer to the fermionic one. However, when a flux $\phi$ threads the ring of electrons the energy is minimum either 
at $\phi = 0$ or $\pi$, depending on whether the number $N$ of electrons on the ring is odd or even, respectively. This shift in 
energy is due to the statistical phase that arises when an electron goes around the ring, thereby changing its place with the other 
$N-1$ electrons. Such 
a parity effect does not arise in the corresponding hardcore bosonic system whose many-body wave function is symmetric under particle 
interchange. Thus, the energy minimum is at $\phi_0=0$ and this is the point about which eq.~\eqref{eq: SFdensity} is evaluated. The 
location of the energy minimum when NNN hopping is included has also been found to be at $\phi_0=0$, provided the system is not in the 
incommensurately ordered phase obtained in Ref. \citen{Mishra}. In Fig.~\ref{fig: Energy} we show a representative plot of the flux dependence of the first three 
energy levels, for a system of two hardcore bosons over a 16-site ring, with $t_2 = 0.2 \, t_1$ and $V=t_1$. The energy crossing at 
$\phi = \pi$ results in a jump discontinuity of the persistent flow, and the energy levels $-$ as well as all physical properties $-$ will 
clearly be periodic functions of $\phi$ with period $2\pi$.

 \begin{figure}
 \centering
 \includegraphics[width=9cm]{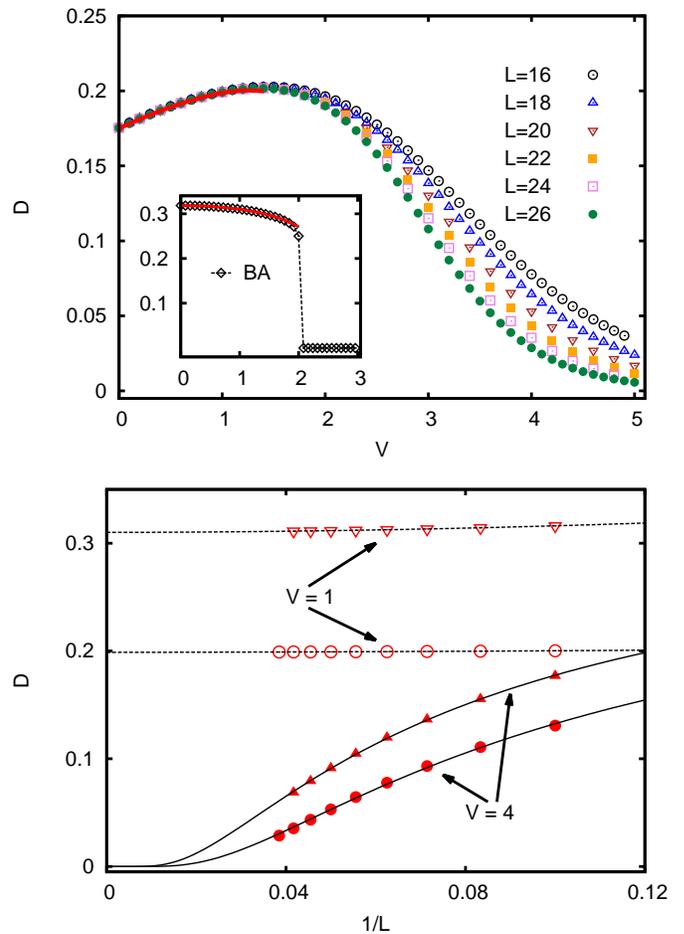}
  \caption{(Colour online) Drude weight of half-filled bosonic rings for various chain lengths $L$. 
  Top panel: Non-monotonicity of the Drude weight with increasing interaction 
  for frustrated nearest-neighbour hopping $t_2= -0.2 t_1$, and polynomial extrapolation (full lines) from data for  
  $L=10, 12, \,  ... \, 26 $ in the conducting phase. Inset shows results from a similar 
  extrapolation for $t_2 = 0$ compared to Bethe Ansatz (BA) results. 
  Bottom panel: Scaling of the Drude weight for $t_2 = 0$ (triangles) and $t_2 = -0.2t_1$ (circles) as a function of inverse system 
  length for interaction strengths $V = 1$ (open symbols) and $V = 4$ (closed symbols). 
  Full lines are exponential fits $D(L) = C\exp{(-L/\xi)}$ and dashed lines are polynomial fits $D(L) = D + a/L + b/L^2$ to 
  the thermodynamic limit $L \rightarrow \infty$. The fits for $V=4$ give a localisation length 
  $\xi \approx 15, 11$ for $t_2 = 0, -0.2t_1$ respectively. As $V$ decreases the fits give increasing estimates for the 
  localisation lengths $\xi$.}
  \label{fig: Extrapolated}
 \end{figure}
  \vspace*{-0.5cm}
\subsection*{Conductor-insulator transition} 

A transition from a conducting to an insulating (charge density wave) phase is expected as the nearest-neighbour repulsive interaction 
strength $V$ is increased, and this is true whether $t_2$ is present or not \cite{Shastry, Mukerjee}{}. In the latter case, the 
Berezinskii-Kosterlitz-Thouless (BKT) transition point is known to be exactly at $V=2 \, t_1$ for a half-filled system. On the other 
hand, in the presence of NNN hopping, the BKT transition points have been computed numerically by Mishra {\it et al.} \cite{Mukerjee} 
from the behaviour of the single particle excitation gap, using the density matrix renormalisation group technique. Here we focus on 
the conducting phase of the system at half-filling and consider the representative frustrated value of $t_2 = -0.2 \, t_1$ for which 
the phase transition was predicted \cite{Mukerjee}{} to occur at $V_c\approx 1.4 \, t_1$. Our main results are summarised in 
Fig.~\ref{fig: Extrapolated} for a range of system lengths. As expected from previous works 
\cite{Kohn,Poilblanc,Millis, Shirakawa}{}, we find the Drude weight to scale polynomially $D(L) = D + a/L + b/L^2$ in the conducting 
phase and exponentially $D(L) = C\exp{(-L/\xi)}$ in the insulating phase, where $\xi$ is the localisation length which diverges at the 
transition, as shown in the bottom panel of Fig.~\ref{fig: Extrapolated}. 

We point out that we do not, nonetheless, employ the scaling laws to identify the transition point. 
Instead, we assume the transition takes place at $V_c = 1.4 \, t_1$ \cite{Mukerjee} and perform a polynomial extrapolation to the 
thermodynamic limit of the computed Drude weights within the conducting phase. Such an extrapolation, shown as the full line in the top panel of 
Fig. \ref{fig: Extrapolated}, works well both for $t_2 = -0.2t_1$ and for the exactly solvable case of $t_2 = 0$ 
(inset of figure, showing agreement of the extrapolation with Bethe Ansatz results \cite{Shastry}). 
For a finite $t_2$ a similar sharp jump in the Drude weight 
should occur as $L \rightarrow \infty$, which expectation is borne out from the plots.

We now expatiate on the singular interplay of interactions and frustration, and its effect on the Drude weight. The most important feature of the 
results shown in Fig.~\ref{fig: Extrapolated} is the rise of the Drude weight as the nearest-neighbour 
interaction is increased, which translates as an enhancement of the persistent flow in the ring. As pointed out in the introduction, an 
interaction-induced enhancement of persistent flows and conducting behaviour has been reported in earlier theoretical studies of 
strongly correlated electronic \cite{Giamarchi, Berkovits, Sarma, Ogata2}{}, hard-\cite{Ogata}{} and soft-core bosonic \cite{Cominotti}{} 
systems, in different \cite{Giamarchi, Berkovits, Sarma} and similar \cite{Ogata, Ogata2, Cominotti} scenarios. Here we uncover a 
mechanism which produces an enhancement of the Drude weight in ladder geometries as a consequence of the competition between non-local 
repulsive interactions and kinetic frustration within the conducting phase, even after a finite-size scaling to the thermodynamic limit 
is performed. Indeed, kinetic frustration favours a situation in which the hardcore bosons gain kinetic energy by hopping back and forth 
between a pair of nearest-neighbour sites and decoupling from the rest of the system \cite{Mukerjee}{}. The repulsive nearest-neighbour 
interaction $V$, on the other hand, favours a charge density wave like ordering where the particles tend to avoid sitting next to each 
other. Thus, initially increasing $V$ causes a spread of the particles and a concomitant increase in charge density wave fluctuations, 
which in turn reduces the effective frustration, thereby increasing the Drude weight. However, upon further increase of $V$ the 
insulating behaviour dominates and the Drude weight drops accordingly. And crucially, as seen in Fig.~\ref{fig: Extrapolated}, such a scenario of 
interaction-induced enhancement of $D$ is valid for finite rings as well as in the thermodynamic limit.

This result is in fact more generic and we have checked its validity away from half-filling. The main difference in that case being a 
much more gradual decay of the Drude weight after it reaches its maximum, which is presumably due to the absence of an insulating phase 
away from half-filling. We have also simulated a system with unfrustrated hopping ($t_2>0$) and found a monotonic decay of $D$ with 
increasing interaction, thereby confirming the crucial role played by the frustration: in the absence of kinetic frustration only one 
localising mechanism is present. 
\vspace*{-0.25cm}
\section{Conclusions}
We have studied the interplay of kinetic frustration with quasi-disorder and interactions on the Drude weight of single-particle and
many-body (quasi) one-dimensional systems, respectively. For the single-particle Aubry-Andr\'{e} model, we found that the Drude weight drops monotonically with an increase 
in the strength of the incommensurate potential and that, in agreement with earlier work \cite{Ramakumar}{}, kinetic frustration allows 
for a quicker onset of the delocalisation-localisation transition.

For the many-body system of nearest-neighbour hardcore interacting bosons, we found that a frustrating nearest neighbour hopping 
causes an interaction-induced enhancement of the Drude weight in the conducting phase. This effect arises because of the degradation of 
the frustration by the nearest-neighbour interaction, causing an increase of the Drude weight; as the non-local repulsive interactions are 
further increased the insulating behaviour sets in and the Drude 
weight drops. Our study on kinetically frustrated hardcore bosons is complementary to an earlier study of a 1D spin system \cite{Ogata} 
where frustrated interactions was seen to destroy the single-particle gap, and a later related work \cite{Ogata2} which showed the 
enhancement of the Drude weight by frustrated interactions in quarter-filled two-dimensional electronic systems.

Our work therefore provides evidence for a unique situation where the conductivity is enhanced in simple kinetically frustrated bosonic systems, 
and opens up the possibility of future cold-atom experiments and theoretical studies wherein the competition between interaction and 
frustration could produce novel effects in the Drude conductivity and persistent flows, even in the simplest of (quasi) one-dimensional 
systems.  In fact, studies and experimental realisations of interacting and frustrated systems in low dimensions are many and varied in 
the present day. Moreover, there still remains much to be gleaned about the additional effect of disorder on the conductivity of these 
kinetically frustrated systems, and on the effects of such a competition in the predicted supersolid phase \cite{Mukerjee}{}.
\vspace*{-0.5cm}
\section{Acknowledgment}
%
One of us (VKV) thanks A. Nersesyan, S. Mandal, and, in particular, S. Pilati for discussions. We are grateful to 
M. M\"{u}ller for helpful comments on an earlier version of the manuscript.
\bibliography{Ref6}

\end{document}